\documentclass[
reprint,
showpacs,
nofootinbib,
 amsmath,amssymb,
 aps,
pra,
floatfix,
]{revtex4-1}

\usepackage{graphicx}% Include figure files
\usepackage{dcolumn}% Align table columns on decimal point
\usepackage{bm}% bold math
\usepackage[colorlinks=true, citecolor=blue, urlcolor=blue ]{hyperref}% add hypertext capabilities
\usepackage{pxfonts,txfonts}
\usepackage{caption}
\usepackage{subcaption}
\newcommand{\CC}{\mathbb{C}}

\newcommand{\HH}{\mathcal{H}}

\newcommand{\bq}{\begin{equation}}
\newcommand{\eq}{\end{equation}}
\newcommand{\bqs}{\begin{equation*}}
\newcommand{\eqs}{\end{equation*}}
\newcommand{\ba}{\begin{array}}
\newcommand{\ea}{\end{array}}
\newcommand{\bas}{\begin{array*}}
\newcommand{\eas}{\end{array*}}
\newcommand{\bqa}{\begin{eqnarray}}
\newcommand{\eqa}{\end{eqnarray}}
\newcommand{\bqas}{\begin{eqnarray*}}
\newcommand{\eqas}{\end{eqnarray*}}

\newcommand{\etal}{\textit{et al.}}

\newcommand{\ran}{\rangle}
\newcommand{\lan}{\langle}

\newtheorem{theorem}{Theorem}

\newtheorem{example}{Example}

\begin{document}

%\preprint{APS/123-QED}

\title{Negative eigenvalues of partial transposition of arbitrary bipartite states}
\author{Swapan Rana}
\email{swapanqic@gmail.com}
\affiliation{Physics and Applied Mathematics Unit, Indian Statistical Institute, 203 B T Road, Kolkata, India}
\date{\today}

\begin{abstract} The partial transposition of a two-qubit state has at most one negative eigenvalue and all the eigenvalues lie in $[-1/2,1]$. In this Brief Report, we extend this result  by Sanpera \etal  [A. Sanpera, R. Tarrach and G. Vidal, \href{http://dx.doi.org/10.1103/PhysRevA.58.826}{Phys. Rev. A {\bf 58}, 826 (1998)}] to arbitrary bipartite states. We show that partial transposition of an $m\otimes n$ state can not have more than $(m-1)(n-1)$ number of negative eigenvalues. Low-dimensional states have been studied to show the tightness of this result and explicit examples have been provided for $mn\le 9$. It is also shown that all the eigenvalues of partial transposition lie within $[-1/2,1]$. Some possible applications are also discussed. 
\end{abstract}


\maketitle

Characterization of entangled states is an important issue in quantum information theory,  from both a theoretical as well as an experimental perspective. Unfortunately, even for the bipartite states, it may be very difficult to decide whether a given state is entangled or not \cite{Gurvits03, GharibianQIC10}. However,  there are still many effective ways to detect entanglement, particularly for low dimensional systems. Undoubtedly, the most useful one is the positive partial transposition (PPT) criteria, introduced by Peres in his seminal work \cite{PeresPRL96}.

For an $m\otimes n$ state $\rho$ acting on the Hilbert space ${\HH}_A\otimes{\HH}_B$, its partial transposition (PT) with respect to the subsystem $A$ is formally defined by $\rho^{T_A}:=(T\otimes I)\rho$, with $T$ being the usual \textit{transposition} map and $\rho^{T_B}$ defined in a similar manner. If $\{|i\ran\}$ is an orthonormal basis of $\HH_A$, then the PT can be computed as \bq\label{firstdefpt} \rho^{T_A}=\sum_{i,j}|j\ran\lan i|\otimes\lan i|\rho|j\ran.\eq Evidently,  $\rho^{T_A}$ depends explicitly on the chosen basis $\{|i\ran\}$, but its eigenvalues do not. So, while considering properties related to eigenvalues, we use $\rho^\Gamma$ to indicate that the result is independent of the chosen subsystem. If $\rho^\Gamma\ge 0$, then $\rho$ is called PPT, otherwise nonpositive partial transposition (NPT). It is well known that separable states are PPT and the converse holds only for $mn\le 6$ \cite{PeresPRL96, MPRHorodeckiPLA97}. Also, NPT states are necessarily entangled and the negativity, a well known measure of mixed state entanglement, is defined as the absolute value of the sum of the negative eigenvalues of $\rho^\Gamma$ \cite{VidalWernerPRA02}. So, the negative eigenvalues of $\rho^\Gamma$ not only certify but also quantify the amount of entanglement in $\rho$. Thus, it is important and interesting to explore the negative eigenvalues of $\rho^\Gamma$.

The two-qubit case has been solved by Sanpera \etal \cite{SanperaetalPRA98} more than a decade ago. It was shown that the PT of a two-qubit state has at most one negative eigenvalue and all the eigenvalues lie within $[-1/2,1]$ (of course, a two qubit state has a negative eigenvalue iff it is NPT and hence entangled). Surprisingly, apart from some conjectures, this beautiful result has not been extended to arbitrary states \cite{AlietalAR07,Xi-JunetalAr06}. Recently we have shown  in Ref. \cite{RanaParasharPRA122} that PT of a $2\otimes n$ state can have at most $(n-1)$-number of negative eigenvalues. Examples of such states are provided in Ref. \cite{ChenDjokovicAR12}. However, the general $m\otimes n$ case is not yet known. In this Brief Report, we will solve this problem. It should be mentioned that based on some numerical findings, the authors of Ref. \cite{Xi-JunetalAr06} have conjectured that the PT of an $n\otimes n$ state can have at most $n(n-1)/2$ number of negative eigenvalues. We will show that contrary to this, the maximum number of negative eigenvalue of PT could go up to $(n-1)^2$. More generally, for an $m\otimes n$ state, we have the following result.

\begin{theorem}
\label{th1} Partial transposition of any $m\otimes n$ state can not have more than $(m-1)(n-1)$ number of negative eigenvalues.
\end{theorem}

\emph{Proof:} Here we will follow a  treatment similar to that of Ref. \cite{SanperaetalPRA98} for the $2\otimes 2$ case. The main ingredient is  Proposition 1.4 from Ref. \cite{ParthasarathyPMS04}, namely, any subspace of dimension $(m-1)(n-1)+1$ of the space $\HH^m\otimes\HH^n$ contains at least one (non-zero) product vector. 

Now, if possible, let the partially transposed state $\rho^{T_A}$ has $(m-1)(n-1)+1$ number of negative eigenvalues $\lambda_i$ with corresponding eigenvectors $|\psi_i\ran$.  Then the hyperplane generated by these $|\psi_i\ran$'s  must contain at least one product vector, say $|e,f\ran$. Therefore, expanding the product vector as $|e,f\ran=\sum c_i|\psi_i\ran$, we get $$\lan e,f|\rho^{T_A}|e,f\ran=\sum_{i=1}^{(m-1)(n-1)+1}\lambda_i|c_i|^2<0.$$ But this would imply $\lan e^\ast, f|\rho|e^\ast,f\ran<0$ which is impossible as $\rho$ is positive semi-definite.\hfill $\blacksquare$

We note that, by Schmidt decomposition, any $m\otimes n$ pure state can be written as \bq\label{pureschmidt}|\psi\ran=\sum_{i=1}^{d\le\min\{m,n\}}\lambda_i|ii\ran,\quad \lambda_i>0,\quad\sum_i\lambda_i^2=1.\eq 
Clearly, its PT is given by $$F:=\sum_{i,j=1}^d\lambda_i\lambda_j|ij\ran\lan ji|.$$ It could be easily checked that $|ii\ran$ and $|ij\ran\pm|ji\ran$ are the eigenvectors of $F$ with the corresponding eigenvalues \bq\label{purePT}\left\{ \begin{array}{ll}
\lambda_i^2, & \forall  i=1,2,\dotsc,d\quad \\
\pm\lambda_i\lambda_j, & \forall  1\le i<j\le d.\quad
\end{array}
\right.\eq 
Thus, for any pure state $P=|\psi\ran\lan\psi|$, its PT, $P^\Gamma$ has $d(d-1)/2$ number of negative eigenvalues. We also observe that, due to the restriction $\sum\lambda_i^2=1$, the following inequality holds, 
\bq\label{purePTineq} -\frac{1}{2}\le\lambda_{\min}(P^\Gamma)\le\lambda_{\max}(P^\Gamma)\le 1.\eq
The bound for $\lambda_{\min}(P^\Gamma)$ could be easily derived using Lagrange's multiplier method, or by setting $x=\lambda_i^2$ and noting that the maximum value of \bq\label{minlambdapure}f(x)=x(1-x-c)\eq over $0\le x\le 1$ and $c\ge 0$ is $1/4$.
The bound for  $\lambda_{\max}(P^\Gamma)$ follows trivially. 

This observation about pure states immediately leads to the following general result.
\begin{theorem}
\label{th3} All eigenvalues of partial transposition of any $m\otimes n$ state always lie within $[-1/2,1]$.
\end{theorem}

\emph{Proof:} Let the spectral decomposition of $\rho$ be given by \bq\label{specrho} \rho=\sum_kp_k|\psi_k\ran\lan\psi_k|:=\sum_kp_kP_k.\eq
Then we have \bqa \lambda_{\min}(\rho^{\Gamma})&\ge&\sum_kp_k\lambda_{\min}(P_k^{\Gamma})\nonumber\\
&\ge&\sum_k p_k.(-\frac{1}{2})\nonumber\\&=&-\frac{1}{2},\label{lammin}\eqa where in the first inequality, we have used the fact that for Hermitian matrices $A_i$, $\lambda_{\min}(\sum A_i)\ge \sum\lambda_{\min}(A_i)$. Similarly, utilizing the dual inequality for $\lambda_{\max}$, we have \bqa \lambda_{\max}(\rho^{\Gamma})&\le&\sum_kp_k\lambda_{\max}(P_k^{\Gamma})\nonumber\\
&\le&\sum_k p_k.1\nonumber\\&=& 1.\label{lammax}\eqa

The tightness of Eq. \eqref{lammin} follows from the fact that partial transposition of the pure state $$|\psi\ran=\sqrt{\frac{1}{2}}|00\ran+\sqrt{\frac{1}{2}-\epsilon}|11\ran+\sqrt{\frac{\epsilon}{m-1}}\sum_{k=2}^m|kk\ran$$ has an eigenvalue $-\sqrt{(1/2)(1/2-\epsilon)}$ where $\epsilon$ could be chosen to vanish. Similarly, the tightness of Eq. \eqref{lammax} follows from the fact that partial transposition of the (separable) state $$\rho=(1-\epsilon)|00\ran\lan 00|+\frac{\epsilon}{m}\sum_{k=1}^m|kk\ran\lan kk|$$ has an eigenvalue $(1-\epsilon)$. Actually, for all pure product states, equality holds in Eq. \eqref{lammax} and no state can saturate both the bounds. \hfill $\blacksquare$

It is clear from Eq. \eqref{pureschmidt} and Eq. \eqref{purePT} that the PT of any $n\otimes n$ pure state with $n$ non-zero Schmidt coefficients will have $n(n-1)/2$ number of negative eigenvalues. This gives the intuition that the maximum number of negative eigenvalues could go beyond the conjectured number $n(n-1)/2$. We will now give several examples to show that this is indeed the case and the bound given in Theorem \ref{th1} is tight.

\begin{example}\label{rhoa}
A  class of $\rho\in \CC^n\otimes\CC^n$ such that $\rho^\Gamma$ has $1+n(n-1)/2$ number of negative eigenvalues.
\end{example} Let us consider the following one parameter family of unnormalized states \bqa\label{1parmnn}\rho_a&=&\sum_{i=1}^3|\psi_i\ran\lan\psi_i|,\\|\psi_i\ran&=&|0i\ran-a|i0\ran,\quad i=1,2,\nonumber\\|\psi_3\ran&=&\sum_{i=0}^{n-1}|ii\ran.\nonumber\eqa
We list the eigenvalues (with multiplicities) of its PT in TABLE-\ref{examplerhoa}. Thus, $\rho_a^\Gamma$ has $n(n-1)/2+1$ number of negative eigenvalues for any $a\in(1/\sqrt{2},1)$. \begin{table}[h]
\caption{\label{examplerhoa}Eigenvalues of $\rho_a^\Gamma$ with multiplicities.}
\begin{ruledtabular}
\begin{tabular}{r|r}
Eigenvalues& Multiplicities\\ \hline
$-1$ &$\frac{n(n-1)}{2}-2$\\
$1$ &$\frac{n(n+1)}{2}-4$\\
$1\pm\sqrt{2}a$&$1$\\
$\frac{1}{2} \left(1+a^2\pm\sqrt{5-2 a^2+a^4}\right)$&$2$
\end{tabular}
\end{ruledtabular}
\end{table}

\begin{example}\label{rho33LinChen}
A  class of $\rho\in \CC^3\otimes\CC^3$ such that $\rho^\Gamma$ has $4$ negative eigenvalues.
\end{example}
We first note that the class of states given by Eq. \eqref{1parmnn} qualifies for the $3\otimes 3$ example. However, for a more constructive example, we will generalize the construction of $2\otimes n$ example, from Ref. \cite{ChenDjokovicAR12}. We consider the following family\bqa\label{33LinGen}\rho(a,b,c)
&=&\sum_{i=1}^3|\psi_i\ran\lan\psi_i|,\\|\psi_1\ran&=&|00\ran+a_1|11\ran+a_2|22\ran\nonumber,\\|\psi_2\ran&=&|01\ran+b_1|12\ran+b_2|20\ran\nonumber,\\|\psi_3\ran&=&|02\ran+c_1|10\ran+c_2|21\ran\nonumber.\eqa
It could be easily checked that the characteristic polynomial of its PT has three factors of the form \begin{multline}\label{char33abc} x^3-(1+a_1^2+b_2^2)x^2+(a_1^2-a_2^2-b_1^2+b_2^2+a_1^2 b_2^2-c_1^2 c_2^2)x\\a_1^2a_2^2+b_1^2 b_2^2+c_1^2 c_2^2-a_1^2 b_2^2-2 a_2 b_1 c_1 c_2=0.\end{multline}
Given that all roots are real, the cubic equation $x^3-p^2x+qx+r=0$ always has a positive root. Furthermore, the conditions $q<0$ and $r<0$ are necessary and sufficient for two negative roots. Thus  we could force one of the three factors to have two negative roots while the other two to have only one. That is there always exists real $a,b,c$ such that $\rho^\Gamma(a,b,c)$ has 4 negative eigenvalues. An example is $a_1=1/4,b_1=b_2=1/3,c_1=1/2,c_2=a_2=1$. Indeed there are infinite number of such states.
\begin{figure}[h]
\includegraphics[width=0.35\textwidth]{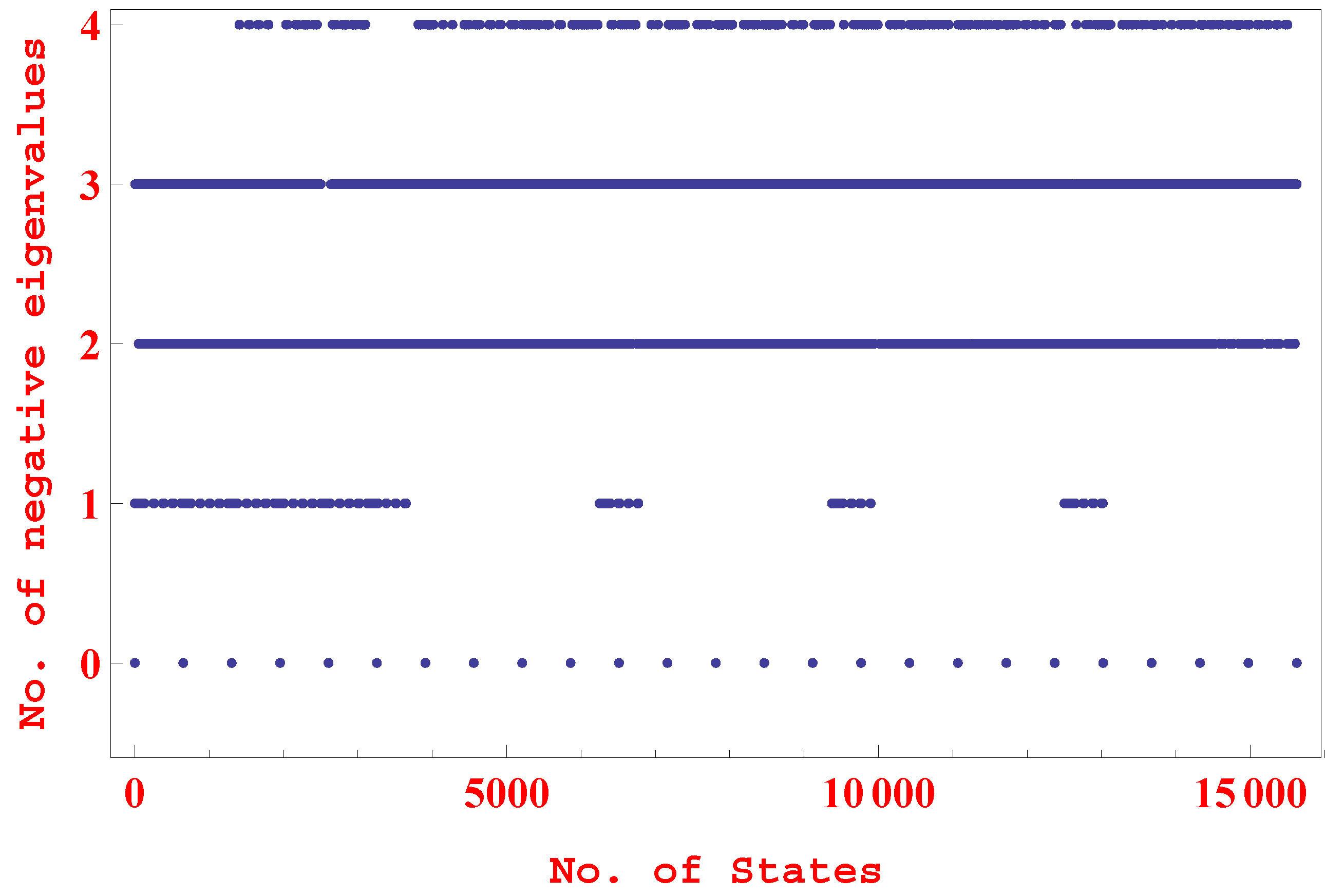}
\caption{\label{fig:33}(Color online ) In  $3\otimes 3$, many $\rho^\Gamma(a,b,c)$, PT of the states in Eq.~\eqref{33LinGen}, have 4 negative eigenvalues. This figure is a \textit{list plot}---each point on the horizontal axis represents a state from the family given by Eq.~\eqref{33LinGen} and the vertical axis represents the number of negative eigenvalues of its PT. For example, the first point corresponds to the state with $a_i=0=b_i=c_i$ and its PT has no negative eigenvalues (see the written text for details).}
\end{figure} 
In FIG.~\ref{fig:33}, we have shown the eigenvalues of $\rho^\Gamma(a,b,c)$  when each of $a_i,b_i,c_i$ takes value from $\{0,1,2,3,4\}/4$. The situation remains almost same, even if we choose $a_i,b_i,c_i$'s as randomly generated complex numbers. 

\begin{example}\label{rho44LinChenFails}
An example of $\rho\in \CC^4\otimes\CC^4$ such that $\rho^\Gamma$ has $8$ negative eigenvalues.
\end{example}
Based on Example \ref{rho33LinChen}, it is tempting to generalize the construction for arbitrary $\CC^n\otimes\CC^n$. We note that the characteristic equation of $\rho^\Gamma(a,b,c,d)$ has 4 factors of the form $x^4-p^2x^3+qx^2+rx+s=0$. In order to $\rho^\Gamma(a,b,c,d)$ have 9 negative eigenvalues, one of the factors must have 3 negative roots and each of the others at least two. The set of constraints thus generated is very complicated for analytic calculations.  We, therefore, have tried to explore numerically and it looks like each such factor has at least 2 positive roots, thereby $\rho^\Gamma(a,b,c,d)$ can not have more than 8 negative eigenvalues. Indeed, there are infinitely many $\rho^\Gamma(a,b,c,d)$ having 8 negative eigenvalues. 
\begin{figure}[h]
\includegraphics[width=0.35\textwidth]{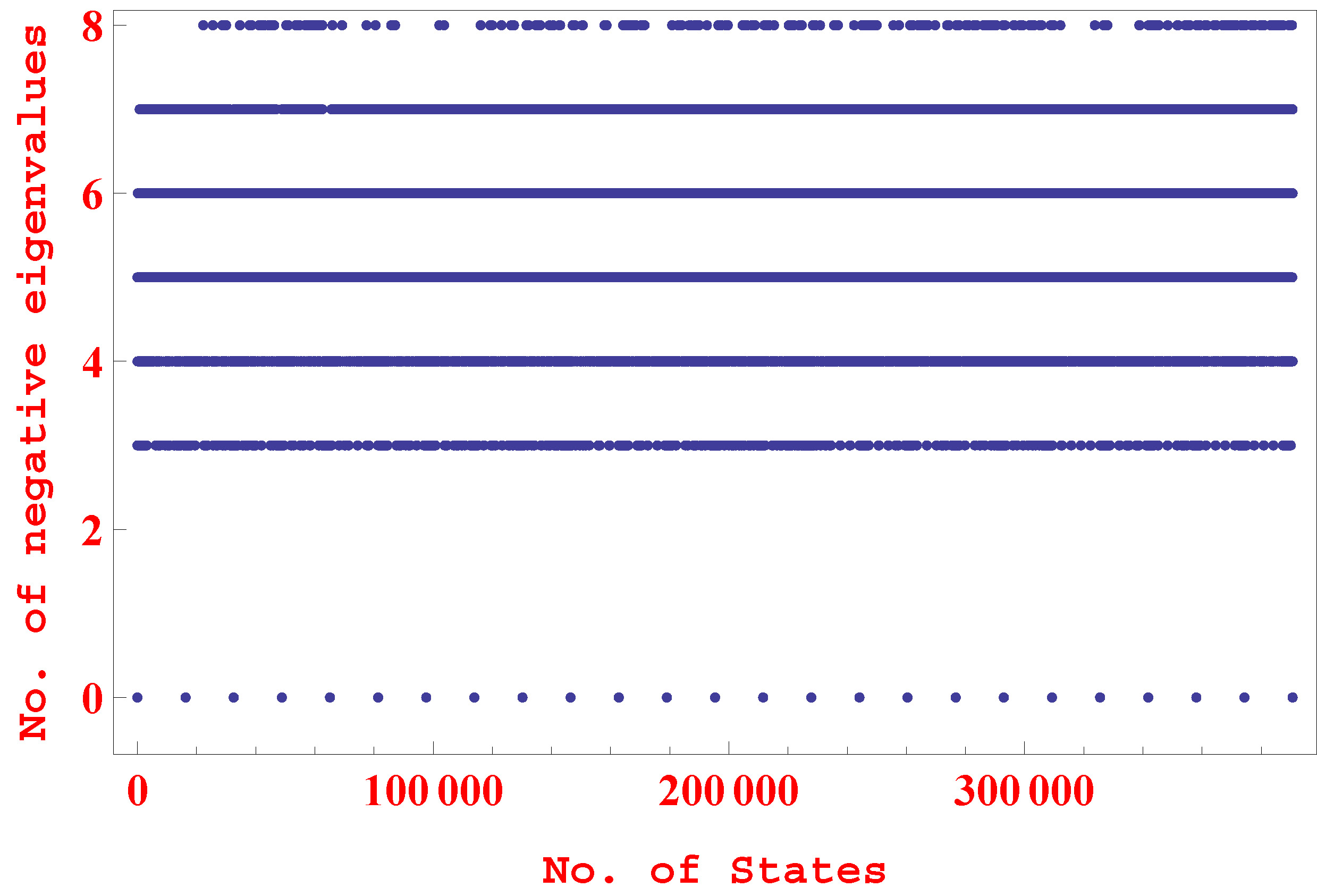}
\caption{\label{fig:44}(Color online) In  $4\otimes 4$, many $\rho^\Gamma(a,b,c,d)$ have 8 negative eigenvalues. However, none seems to achieve the maximum number of negative eigenvalues (nine).}
\end{figure} 
In FIG.~\ref{fig:44} we show some of such states where $a_i,b_i,c_i,d_i$ takes value from $\{2,4,6,8,10\}/10$. Like the previous case, the parameters could be taken as random complex numbers as well. 

We have explored (both numerically and analytically) other small-dimensional states as well. Unfortunately, however, we are unable to settle the question of tightness of the bound $(m-1)(n-1)$, beyond two qutrits. In TABLE~\ref{finaltable} we summarize our findings. \begin{table}[h]
\caption{\label{finaltable}The bound of Theorem~\ref{th1} and its tightness.}
\begin{ruledtabular}
\begin{tabular}{l|l|l}
Dimensions& The bound $(m-1)(n-1)$& maximum achieved\\ \hline
$2\otimes n$ & $n-1$&$n-1$\\
$3\otimes 3$ & 4 & 4\\
$3\otimes 4$ & 6 & 5\\
$3\otimes 5$ & 8 & 6\\
$4\otimes 4$ & 9 & 8
\end{tabular}
\end{ruledtabular}
\end{table}

As mentioned earlier, the main ingredient in the proof of Theorem-\ref{th1} was the result of maximal dimension of entangled subspace from Ref. \cite{ParthasarathyPMS04} and thus the proof is not constructive. However, it appears that the problem could be solved completely using only matrix theoretic techniques. But, the question about the tightness of the bound is yet to be explored.

Although the main motivation for this study was the curiosity of extending the result of two qubits to arbitrary states, nonetheless let us mention some possible applications of this upper bound. Indeed prior to this work, the exact number of negative eigenvalues of PT were applied to get interesting results about small dimensional system. For example, the result of $2\otimes 2$ system has been used to show that all separable states can be expressed as a mixture of at most 4 pure product states. The 2-qubit case being so special, this result, coupled with the fact that NPT is equivalent to full rank of PT, implies that a $2\otimes 2$ state $\rho$ is entangled iff $\det\rho^\Gamma<0$. Clearly, this condition, though always sufficient, is not necessary for separability beyond two-qubits. The pure state after Eq.~\eqref{lammax} also shows that to estimate negativity, not the number of negative eigenvalues of PT, but rather the the one with maximum modulus is significant. Thus, apparently this generic bound, contrary to small-dimensional systems may have less direct physical significance for higher dimensional systems.

Apart from its close connection with the maximal dimension of \textit{completely entangled subspaces} \cite{ParthasarathyPMS04,CubittETALJMP08}, the present bound for arbitrary bipartite states, albeit mostly a mathematical result, may also have some possible applications in quantum information theory. For example, similar to Ref.~\cite{RanaParasharPRA122}, this bound could readily be applied to give a semi analytical proof that squared negativity may exceed geometric discord in higher dimensional states as well and the number of such states will increase with the dimension. It is mentioned in Ref.~\cite{RanaParasharPRA122} that due to lack of knowledge about this generic bound (and also lack of analytic formula for geometric discord), only $2\otimes n$ states were considered. In view of the bound derived here, the said result (about geometric discord and negativity) can be easily arrived at, by following exactly the proof of Theorem 2 therein. For unnecessary repetitions, we skip the details. In another direction, following Ref.~\cite{AlietalAR07}, the result may have some applications in the study of the dynamics of entanglement. 

To conclude, extending a decade old result for two-qubits, we have shown that the partial transposition of a generic $m\otimes n$ state can not have more than $(m-1)(n-1)$ number of negative eigenvalues. Besides giving some explicit examples of tightness in small dimension, we have shown that all the eigenvalues always lie within $[-1/2,1]$. Some consequences of this bound have been discussed, in particular, two possible applications of the results have been mentioned. However, the question of tightness of this bound beyond two-qutrits remains open.

\textit{Note added in the proof.} Recently we found a work \cite{...SMFeiIJTP13} describing an interpretation of the number of 
negative eigenvalues of $\rho^\Gamma$. It has been shown that if for any mixed state $\rho$, $\rho^\Gamma$ has $K+1$ number of negative eigenvalues ($K\ge 1$), then for any $K$ product state $|\psi_k\ran$, the state $\rho+\sum_{k=1}^K\lambda_k|\psi_k\ran\lan\psi_k|,\lambda_k\ne 0$ will always remain NPT. Addition of one more pure product state to $\rho$ may lead to PPT (both separable and entangled) as well as NPT states. 

I am thankful to Guifr\'{e} Vidal for important input. I would also like to thank  Preeti Parashar and Lin Chen for many helpful discussions.


\begin{thebibliography}{50}
\bibitem{Gurvits03} L. Gurvits, in \textit{Proceedings of the 35th ACM Symposium on Theory of Computing} (ACM Press, New York, 2003), 
\href{http://dx.doi.org/10.1145/780542.780545}{pp. 10-19}. 

\bibitem{GharibianQIC10} S. Gharibian,Quantum Inf. Comput. {\bf 10}, 343 (2010). 

\bibitem{PeresPRL96} A. Peres, \href{http://dx.doi.org/10.1103/PhysRevLett.77.1413}{Phys. Rev. Lett. {\bf 77}, 1413 (1996).}

\bibitem{MPRHorodeckiPLA97} M. Horodecki, P. Horodecki, and R. Horodecki, \href{http://dx.doi.org/10.1016/S0375-9601(96)00706-2}{Phys. Lett. A {\bf 223}, 1 (1996).}

\bibitem{VidalWernerPRA02} G. Vidal and R. F. Werner, \href{http://dx.doi.org/10.1103/PhysRevA.65.032314}{Phys. Rev. A {\bf 65}, 032314 (2002).}

\bibitem{SanperaetalPRA98} A. Sanpera, R. Tarrach and G. Vidal, \href{http://dx.doi.org/10.1103/PhysRevA.58.826}{Phys. Rev. A {\bf 58}, 826 (1998).}

\bibitem{AlietalAR07} M. Ali, A. R. P. Rau and K. Ranade, \href{http://arxiv.org/abs/0710.2238}{arXiv:0710.2238}

\bibitem{Xi-JunetalAr06} R. Xi-Jun, H. Yong-Jian, W. Yu-Chun and G. Guang-Can, \href{http://dx.doi.org/10.1088/0256-307X/25/1/010}{Chin. Phys. Lett. {\bf 35}, 35 (2008).} Also available at \href{http://arxiv.org/abs/quant-ph/0609091}{arXiv:0609091}.

\bibitem{RanaParasharPRA122} S. Rana and P. Parashar, \href{http://dx.doi.org/10.1103/PhysRevA.86.030302}{Phys. Rev. A {\bf 86}, 030302(R) (2012).}

\bibitem{ChenDjokovicAR12} L. Chen and D. \v{Z}. DJokovi\'{c}, \href{http://dx.doi.org/10.1103/PhysRevA.86.062332}{Phys. Rev. A {\bf 86}, 062332 (2012).}

\bibitem{ParthasarathyPMS04} K. R. Parthasarathy, \href{http://dx.doi.org/10.1007/BF02829441}{Proc. Indian. Acad. Sci. (Math. Sci.) {\bf 114}, 365 (2004).} Also available at \href{http://arxiv.org/abs/quant-ph/0405077}{arXiv:0405077}. 

\bibitem{CubittETALJMP08} T. Cubitt, A. Montanaro, and A. Winter, \href{http://dx.doi.org/10.1063/1.2862998}{J. Math. Phys. \textbf{49}, 022107 (2008).}

\bibitem{...SMFeiIJTP13} B. Hua, X.-H. Gao, M.-J. Zhao and S.-M. Fei, \href{http://dx.doi.org/10.1007/s10773-013-1591-6}{Int. J. Theor. Phys. doi 10.1007/s10773-013-1591-6 (2013).}

\end{thebibliography}
\end{document}